\documentclass[conference,letterpaper]{IEEEtran_nssmic}
\usepackage[dvips]{graphicx}
\usepackage{cite}
\usepackage{subfigure}
\usepackage{epsfig}

\begin{document}
%
\title{Fluctuation studies and energy reconstruction in a segmented calorimeter}

\author{\authorblockN{Sara~Bergenius~Gavler\authorrefmark{1}, Per~Carlson\authorrefmark{1} and~Jan~Conrad\authorrefmark{1}}\\
\authorblockA{\authorrefmark{1}School of Engineering Sciences\\
Royal Institute of Technology (KTH),
Stockholm, Sweden\\ Email: conrad@particle.kth.se}}

\maketitle

\begin{abstract}
In order to better understand energy estimation of electromagnetic showers in segmented calorimeters, detailed Geant4 simulation studies of electromagnetic showers in the energy range 1--100~GeV in CsI have been performed. 
When sampled in layers of 1.99~cm thickness, corresponding to 1.08 radiation lengths, the energy fluctuations in the samples show distributions that vary significantly with depth.
The energy distributions are similar for incident electrons and photons and were found to change systematically along the shower, varying little with the initial energy.\\
Three probability distributions have been fitted to the data: negative binomial, log-normal and Gaussian distributions, none of which gives a good fit over the full shower length. The obtained parameterizations might be useful in the use of the maximum likelihood method for estimating the energy.\\
Two methods for estimating the energy have also been studied. One method is based on fitting individual longitudinal shower profiles with a $\Gamma$-function, the other one corrects the measured energy for leakage by utilizing the energy deposited in the last layer. Our simulations indicate that the last-layer correction method applied to photons and electrons of 1 and 10 GeV gives about a factor of 2 improvement in the energy resolution.

\end{abstract}


\section{Introduction}
In a segmented calorimeter, where the energy deposition is sampled in layers along the shower, the fluctuations in a layer at a given depth arise from fluctuations in starting point and development of the shower.
Already in 1937 Bhabha and Heitler~\cite{bahba}, in their work on electromagnetic cascades, studied fluctuations and proposed that these follow a Poisson distribution.
Furry~\cite{furry} used a multiplicative model for the shower development and arrived at a wider distribution, often called the Furry distribution.
Mitra~\cite{mitra} used the negative binomial or P\'olya distribution to describe air showers, with the Poisson and Furry distributions as limiting cases.
The P\'olya distribution is mentioned in conjunction with cosmic rays for the first time in 1943~\cite{arley}. It has been used e.g. to describe photomultiplier electron statistics~\cite{presc} and in particle physics to describe hadronic inelastic non-diffractive multiplicity distributions of produced particles~\cite{giovanni,alner}. 
Fluctuations in electromagnetic cascades have been considered as an application of Markov processes~\cite{bharreid} and analytic solutions have been found~\cite{messel}.

Only few experimental or simulation based studies of fluctuations at a given depth can be found in the literature. Longo and Sestili~\cite{longosest} have used simulations to study fluctuations in the last radiation length of lead glass of different lengths for photons with energies up to 1~GeV. They used the P\'olya distribution for comparison.

Gaisser~\cite{gaiss}, in discussing cosmic air showers, points out that these fluctuations are smallest near shower maximum and that they approximately follow a log-normal distribution, reflecting the multiplicative nature of the shower development.
In a recent paper, Souza et al.~\cite{souza} studied the fluctuations of maximum number of particles in air showers in order to estimate how well the shower size at maximum depth can determine the shower energy.
General properties of calorimeter fluctuations are reported e.g. by Amaldi~\cite{amaldi}.
In this contribution we report on a simulation study of fluctuations as function of depth in a segmented CsI calorimeter. We also use these simulations to compare different methods for energy reconstruction especially for non-contained electro-magnetic showers.

%
%
%
%





%

\section{Simulations}
\label{sec:calosim}
\begin{figure}
\centering
\includegraphics[width=2.5in]{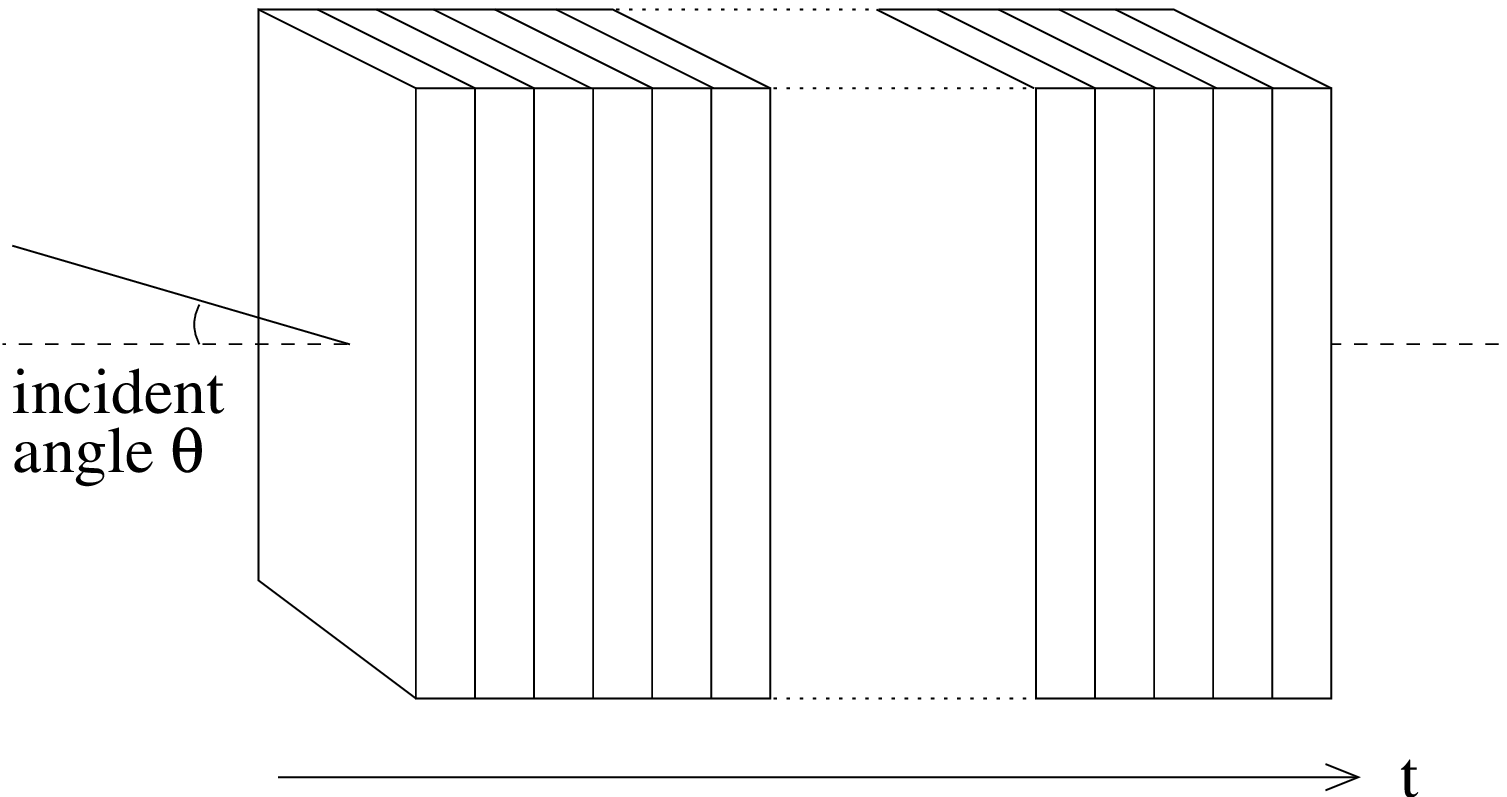}
\caption{The segmented CsI calorimeter used in the simulations. A calorimeter layer consists of 6 crystal bars and the layer number increases with $x$. The incident particle enters the calorimeter from left}
\label{fig:simcal}
\end{figure}

In the simulation studies the calorimeter consisted of CsI (radiation length 1.85~cm) crystal bars arranged in layers perpendicular to the incident particles.
The layout, shown in fig.~\ref{fig:simcal}, resembles a GLAST electromagnetic calorimeter tower~\cite{john}, which is built up by 8 orthogonally oriented layers of 12 CsI(Tl) crystal bars of size 32.60$\times$2.67$\times$1.99~cm$^3$, giving a total thickness of about 8.6 radiation lengths for perpendicular incidence.
In order to study the basic fluctuation phenomena, the simulated calorimeter does not contain any supporting material or gaps, the layers were not placed orthogonal, and in total 20 layers were implemented.
Each layer consisted of 6 crystals.
Furthermore, in the simulations the energy deposited in the crystals was taken as the measured energy, i.e. no readout system has been applied.

The simulations were performed using Geant4 (v7.0p01), a Monte Carlo particle transport toolkit which is well supported and used by several experiments.
Only standard electromagnetic processes were considered, all photo-nuclear interactions have been neglected.
Ajimura et al.~\cite{ajimura} have measured the photo-nuclear interactions to contribute to less than $2\times 10^{-7}$ in CsI for energies above 1~GeV.
Electromagnetic processes are well simulated by Geant4 in the energy range relevant here, for a validation see e.g.~\cite{amako}.
Geant4 uses the particle range to set the energy cut-off below which secondary particles are no longer tracked and the energy is simply dumped at that location. 
We used a range cut of 1~mm, corresponding to an energy cut-off of 38~keV, 692~keV and 658~keV for respectively photons, electrons and positrons in CsI. 
Simulations were performed with incident electrons and photons in the energy range 1--100~GeV with a minimum of 100000 events in each simulation.
Most studies were done with perpendicular input, but the case with non-perpendicular input having a incident angle of $30^\circ$ was also examined.

%

%

\section{Fluctuation studies}

\label{sec:results}

\subsection{Longitudinal profiles}
\begin{figure}
\centering
\includegraphics[width=2.5in]{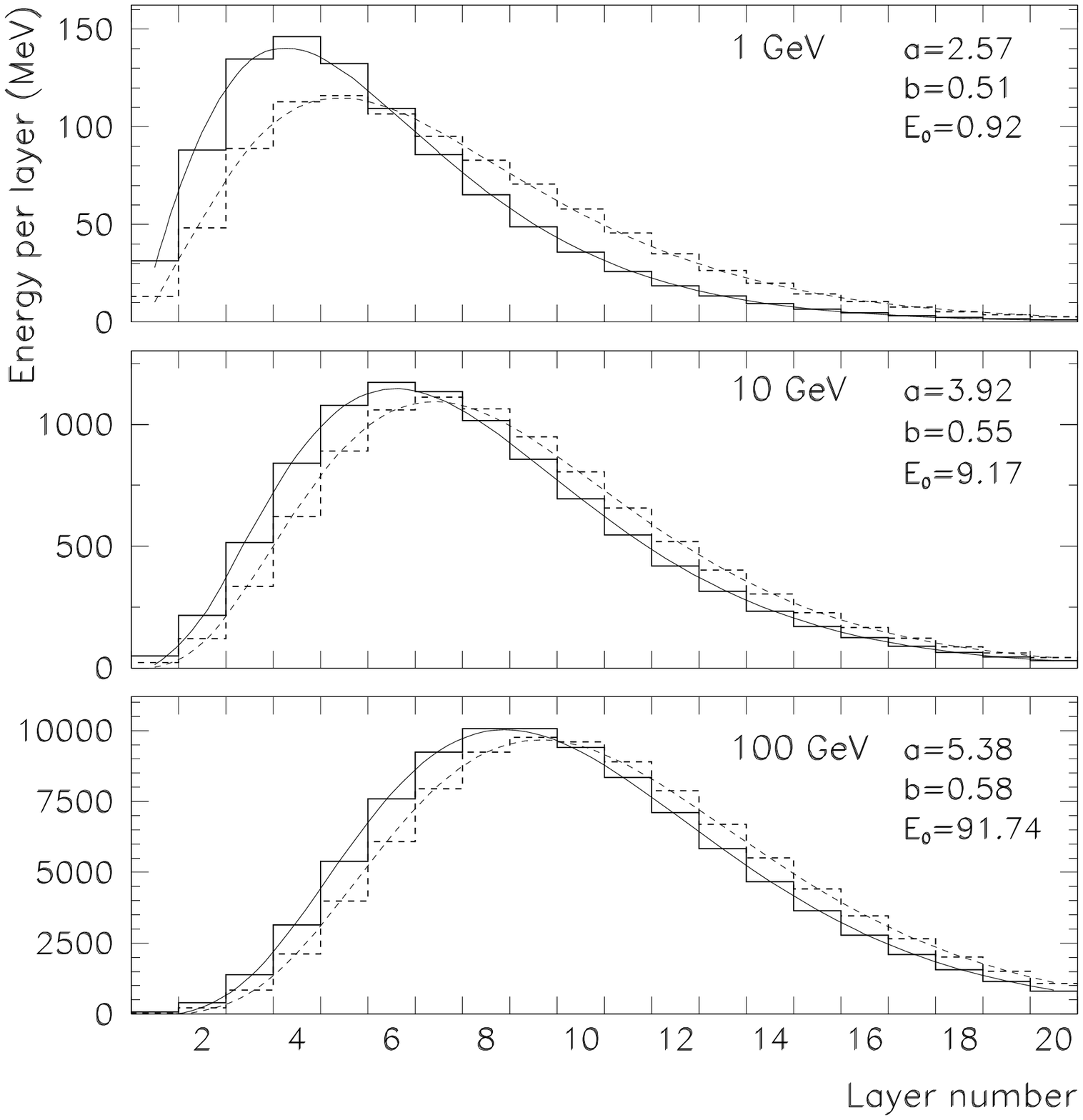}
\caption{Mean energy deposition in the calorimeter layers, here shown for both electrons (solid line) and photons (dashed line) having initial energies of 1~GeV, 10~GeV and 100~GeV. The distributions have been fitted with a Gamma function (eqn.~\ref{eqn:gamma}).}
\label{fig:meanedep}
\end{figure}

Fig.~\ref{fig:meanedep} shows the mean longitudinal profiles for showers induced by electrons and photons having initial energies of 1~GeV, 10~GeV and 100~GeV.
The showers induced by photons reach their maxima later than those induced by electrons.
This is due to the differences in electron and photon interaction mechanisms.
An electron starts losing energy immediately as it enters the calorimeter media, while photons may travel a distance before interacting.
See further the work of Wigmans and Zeyrek~\cite{wigmans} for a study on differences between photon- and electron-induced showers. The profiles are well fitted with the Gamma distribution:
\begin{equation}
\frac{dE}{dt} = E_{0}b\frac{(bt)^{a-1} e^{-bt}}{\Gamma(a)}
\label{eqn:gamma}
\end{equation}
where $t$ is the depth in radiation lengths, $a$ an energy-dependent parameter and $b$ a $Z$-dependent parameter varying slowly with energy.
$E_0$ is the initial energy of the particle, $\Gamma$ is the Gamma function.

We see that the shower maxima for initial energies of 1~GeV, 10~GeV and 100~GeV occur at layer 4, 6 and 8 (4.3, 6.5 and 8.6 radiation lengths) for electrons and at layer 5, 7 and 9 (5.4, 7.5 and 9.7 radiation lengths) for photons.

\subsection{Sample fluctuations}
\begin{figure}[t]
\centering
\includegraphics*[width=2.5in,clip=true]{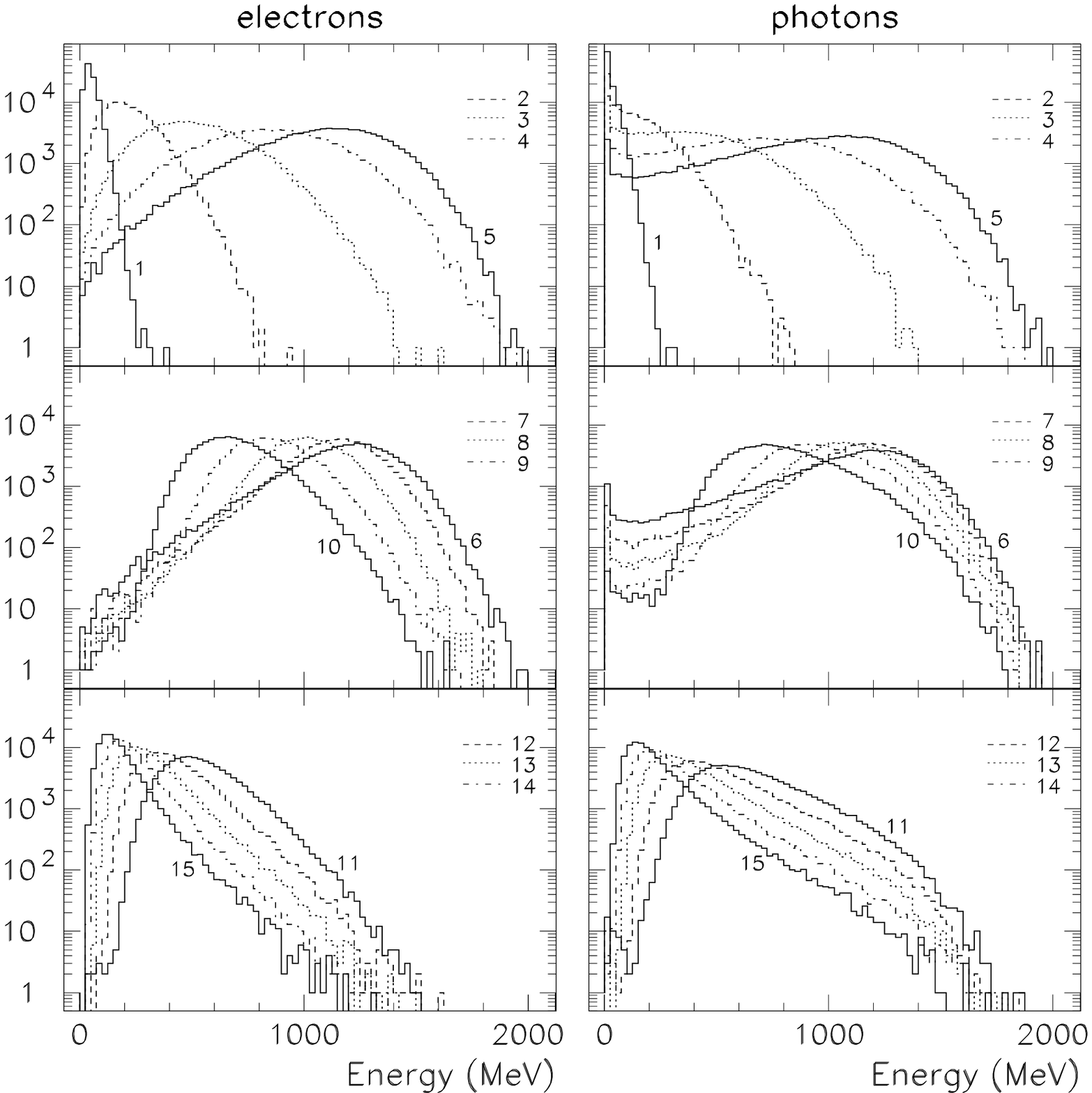}
\caption{Fluctuations of energy deposition in the CsI calorimeter layers for showers induced by 1~GeV electrons and photons. The histograms show the number of events depositing in an energy interval corresponding to the bin size (here 4~MeV). Each layer of CsI is 1.99~cm (1.08 radiation lengths).}
\label{fig:fluct_lay}
\end{figure}

The fluctuations in energy deposition in each layer were studied both for showers induced by electrons and photons of initial energy of 1~GeV, 10~GeV and 100~GeV.
In each case 100000 incident particles (electrons or photons) were simulated.
Figs.~\ref{fig:fluct_lay} shows how the fluctuations change from layer to layer in the first 15 layers (16 radiation lengths) for a particle energy of  10~GeV.The energy fluctuations in layer 16--20 have the same exponential-like shape as layer~15, only shifted towards lower energy as the depth increases.
All of the 10~GeV electrons have deposited energy in the first layer (1.1 radiation lengths) whereas about 5\% of the 10~GeV photons have not interacted at this depth.

%

The shape of the energy fluctuations changes systematically with depth for the three different energies.
Both electron- and photon-induced showers show fluctuations with a high-energy tail at small depths, which is reduced as the depth increases and the distribution becomes more symmetric and Gaussian-like.
Moving deeper into the calorimeter, a low-energy tail develops.
Eventually the fluctuations become more symmetric again and finally a high-energy tail emerges.
The non-asymmetry of the energy distributions can be readily understood from fluctuations in the shower developments.
Indeed, in the first few radiation lengths early shower developments can give rise to large total energy deposits giving a high-energy tail.
At the location of the mean maximum energy deposition, fluctuations to higher energies are less frequent than those to lower energies, giving distributions with a slight low-energy tail.
The most symmetric Gaussian-like distributions occur just before and after the maximum energy deposition where there are both high- and low-energy contributions from early respectively late initiated showers.
These tendencies become more pronounced with increasing energy as the longitudinal shower profile becomes more stretched out.


The energy fluctuations in each layer were fitted with negative binomial, log-normal and Gaussian distributions for incident electrons and photons of energy 1~GeV, 10~GeV and 100~GeV.
The fits were performed over a range in energy corresponding to 1\% of the peak value of the distribution.
As expected from the variations as function of depth, quality of fits varies considerably and in most cases there are deviations from the fitted form in the tails of the distributions. An important conclusion from the fits is that for each layer one -- but not the same -- distribution gives the best fit.
With 100000 simulated events only a few layers give a fully acceptable fit as judged by the $\chi^2$.

\begin{figure}
\centering
\includegraphics[width=2.5in]{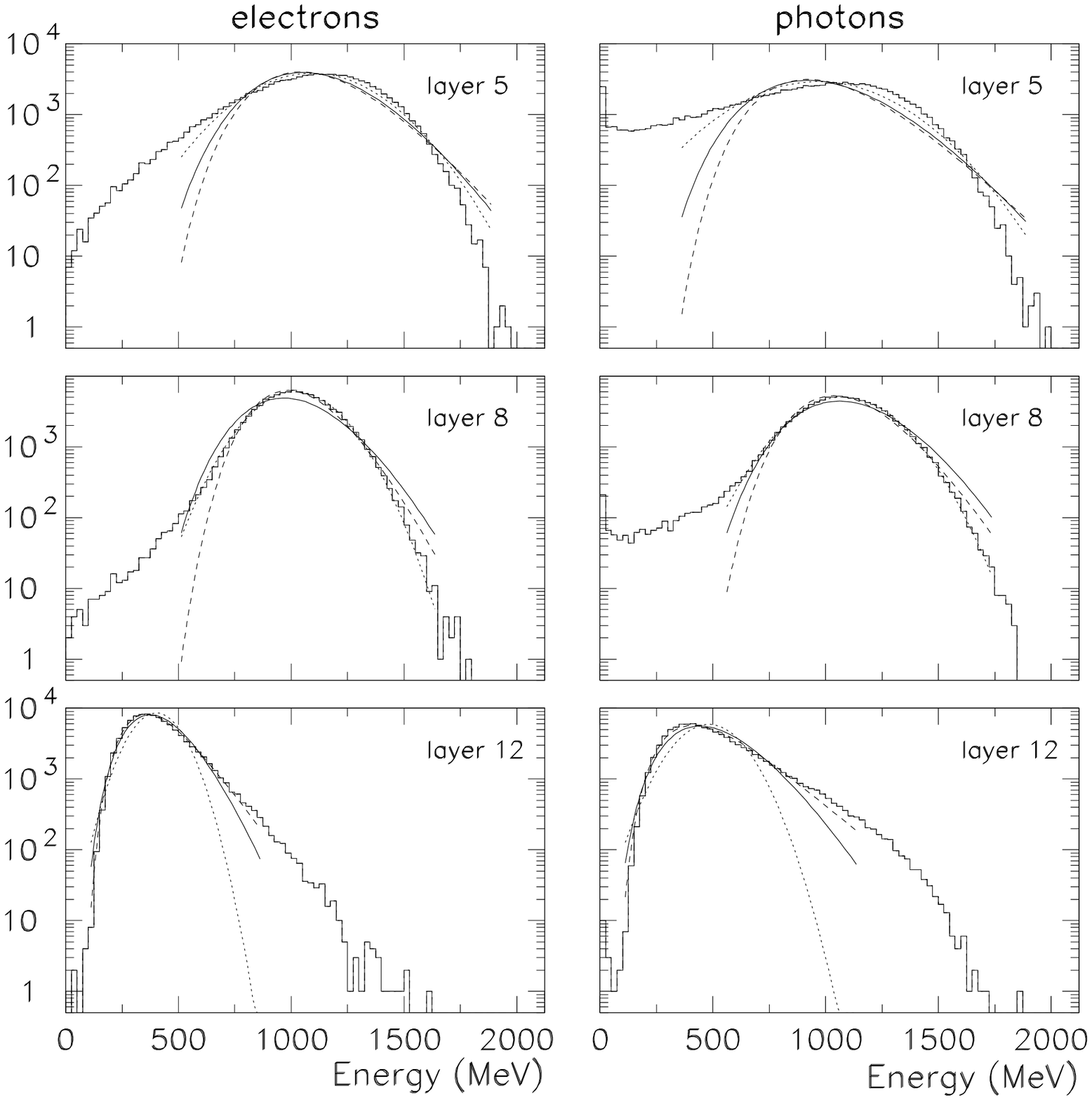}
\caption{Fitting the energy fluctuations for incident electrons and photons having an initial energy of 10~GeV with negative binomial (solid line), log-normal (dashed line) and Gaussian (dotted line) distributions. The histograms show the number of events depositing in an energy interval corresponding to the bin size (here 25~MeV). The fitting range covers 97.1, 99.5 and 98.9\% (88.8, 97.8 and 98.4\%) of the distribution for electrons (photons).}
\label{fig:fluctfit}
\end{figure}

As an example we show in fig.~\ref{fig:fluctfit} the 10~GeV electron and photon energy distributions in layers 5, 8 and 12 (5.4, 8.6 and 12.8 radiation lengths). These three layers are before, close to and after the shower maximum.
The fitting range covers 97.1, 99.5 and 98.9\% (88.8, 97.8 and 98.4\%) of the 100000 events in the simulation for electrons (photons).

\begin{figure}
\centering
\includegraphics[width=2.5in]{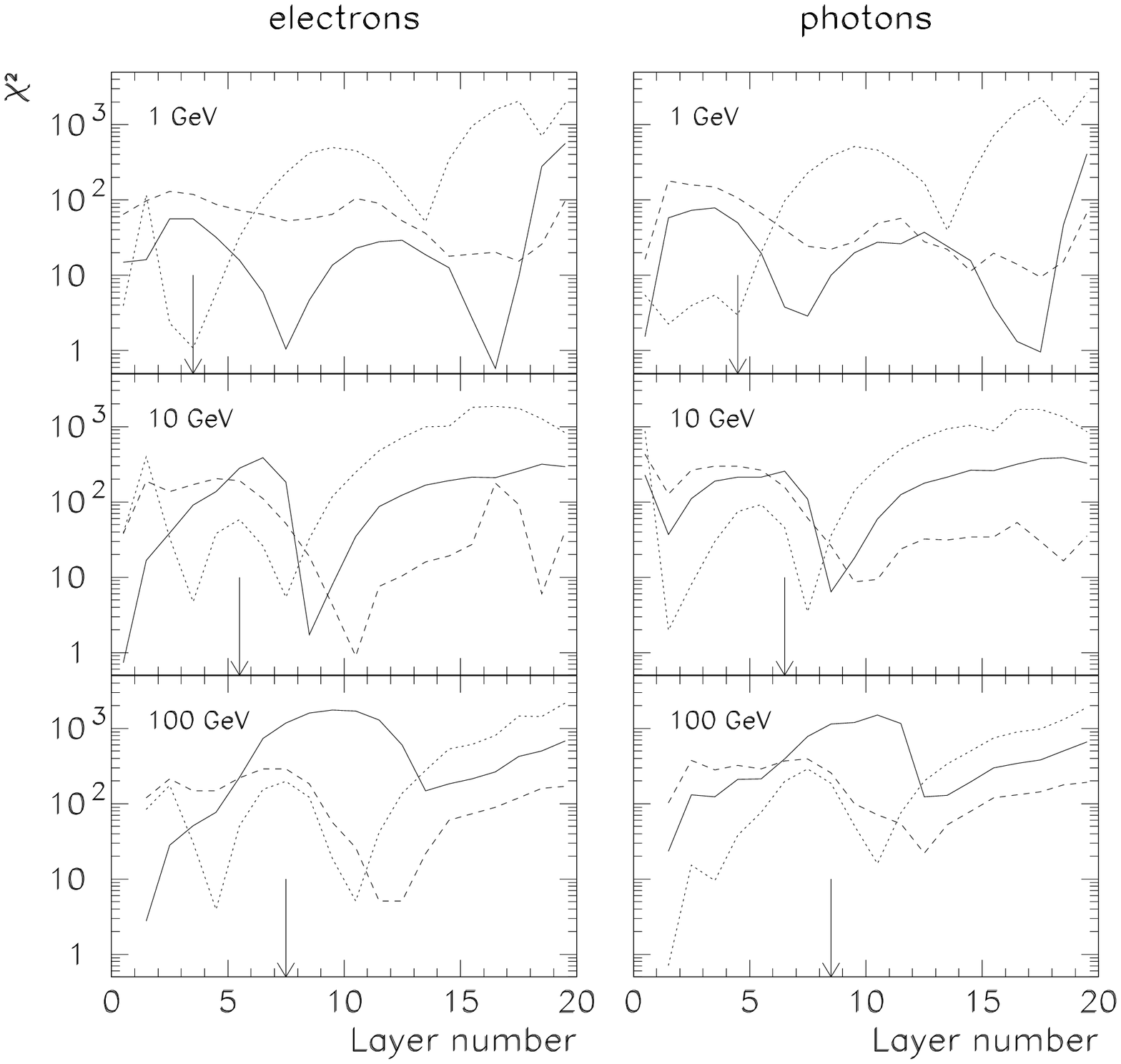}
\caption{The reduced $\chi ^2$ of the negative binomial (solid line), log-normal (dashed line) and the Gaussian (dotted line) fit for 1~GeV, 10~GeV and 100~GeV electrons and photons. The lines have been drawn to guide the eye. The arrows mark the mean shower maximum.}
\label{fig:parchi}
\end{figure}

The low-energy tails in layers 5 and 8 are not well fitted. The high-energy tails are quite well described by the log-normal or negative binomial distributions.
Fig.~\ref{fig:parchi} shows as function of layer number the reduced $\chi^2$ of the fits, giving a relative evaluation of how well the three distributions fit the energy fluctuations. The systematic change of the fluctuations with depth can be seen.
We include in fig.~\ref{fig:parchi} the reduced $\chi^2$ for layers 16--20 although the number of events there is small. For 1~GeV incident electrons and photons, the Gaussian distribution gives the best fit in the first 5 layers which is expected since the the mean shower maximum occurs in layer 4. At larger depths the energy fluctuations are best fitted with the negative binomial. Note that the energy fluctuations become close to exponential at layer 14 and beyond, giving a large uncertainty in the fitted parameters.
For 10~GeV the Gaussian distribution gives the best fit a few layers before and after the mean shower maximum, as expected.
When the high-energy tail emerges, the best fit is given by the negative binomial, and as it becomes more prominent, the log-normal.
For 100~GeV the Gaussian again gives the best fit before and after the location of the shower maximum, and when the high-energy tail emerges the log-normal distribution gives the best fit.
The negative binomial gives quite a good fit up to layer 4, but after that it completely fails.
There are no larger differences in the $\chi^2$ between electron- and photon-induced showers since the width changes but not the shape of the fluctuations.

In this way parameterizations of the deposited energy distributions can be obtained as function of energy and incident angle.
More detailed results of this simulation study are presented in~\cite{sarabe}.

%
%

\section{Comparison of energy reconstruction methods}

\subsection{The method of last layer correction.}

\begin{figure}
\centering
\includegraphics[width=2.5in]{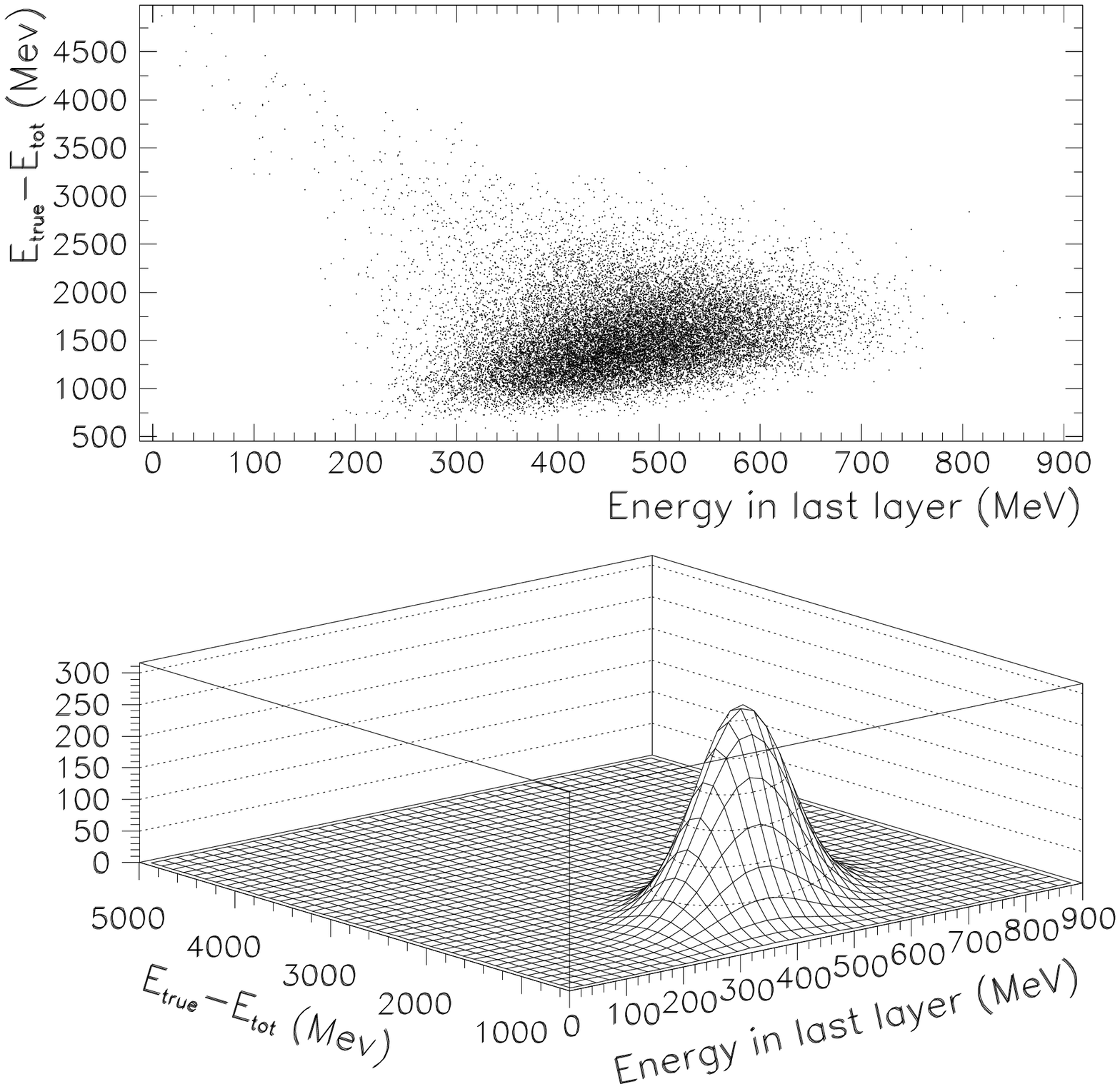}
\caption{An example for the correlation between energy deposited in the last layer and the energy lost due to leackage. The lower panel is a three dimensional presentation of the upper panel.}
\label{fig:last_layer}
\end{figure}

The simplest method to estimate the energy in a segmented calorimeter is to sum up the energies deposited in each segment. This estimate is only accurate if the energies of the incident particles are small as compared to the average energy lost of the particles in the calorimeter. In the usual case, corrections have to be applied for the energy lost. In this context it is worth noting that the energy lost due to leakage is correlated with the energy deposited in the last layer (see figure \ref{fig:last_layer}). A corrected energy can thus be calculated from:
\begin{equation}
E_{cor} = E_{tot}+kE_{last}
\end{equation} 
where $E_{cor}$ is the energy estimated corrected for leakage, $E_{tot}$ is the total energy deposited in the calorimeter and 
$E_{last}$ is the energy deposited in the last layer. The parameter $k$ is to be determined from Monte-Carlo simulations or from measurements under controlled conditions (beam test).

\subsection{Fitting of individual profiles with the $\Gamma$- distribution.}
\begin{figure}
\centering
\includegraphics[width=2.5in]{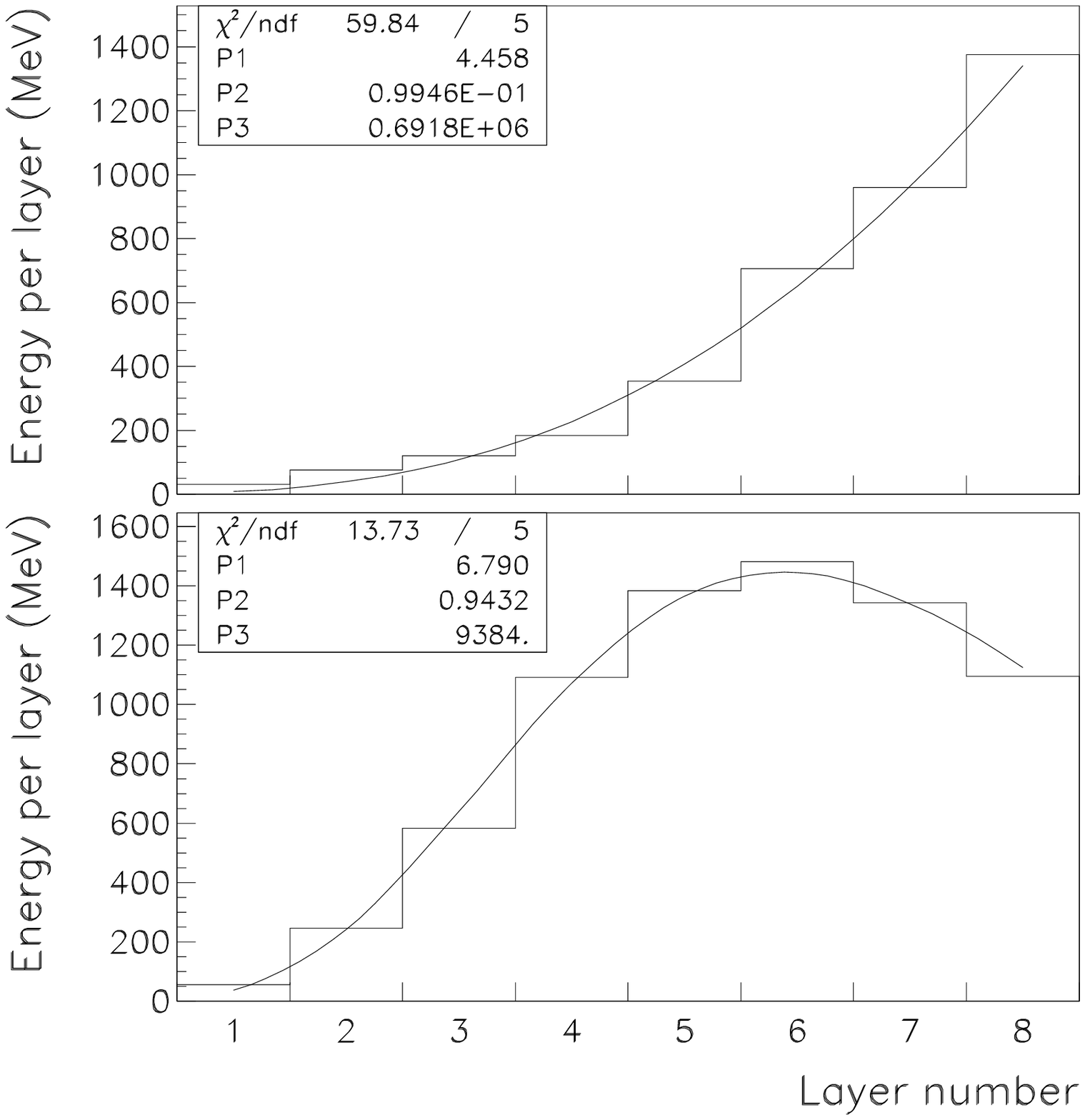}
\caption{Two examples of individual longitudinal shower profiles fit with the $\Gamma$-function. In this case, 10 GeV photons have been simulated at normal incidence. The upper panel shows a longitudinal profile with its maximum outside the sensitive calorimeter volume. The energy estimate obtained in this case is 692 GeV. The lower panel shows an example of a profile where the shower maximum is contained. The estimated energy in this case is 9.4 GeV.}
\label{fig:badindfit}
\end{figure}

\begin{figure}
\centering
\includegraphics*[width=2.5in,clip=true]{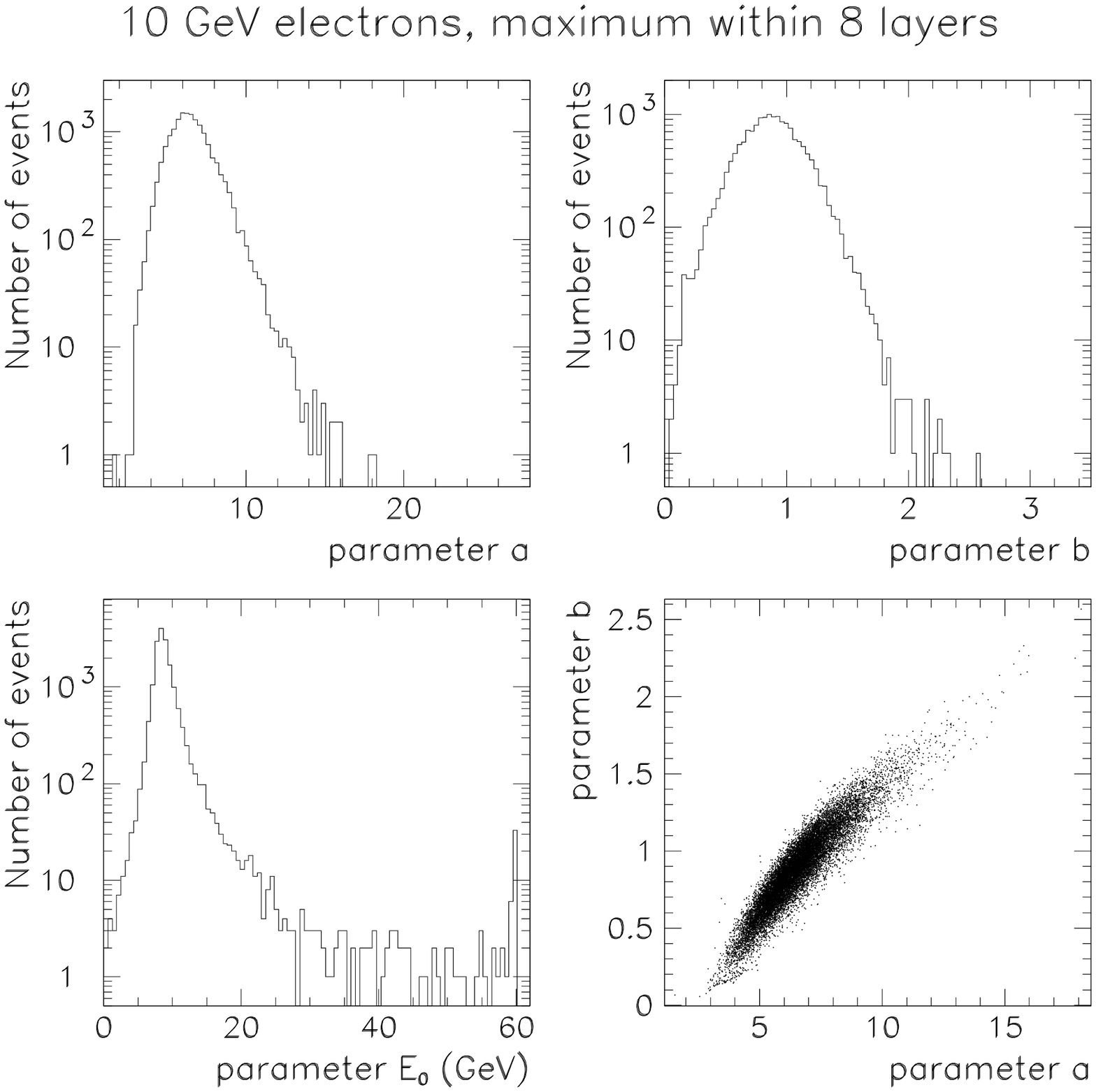}
\includegraphics[width=2.5in]{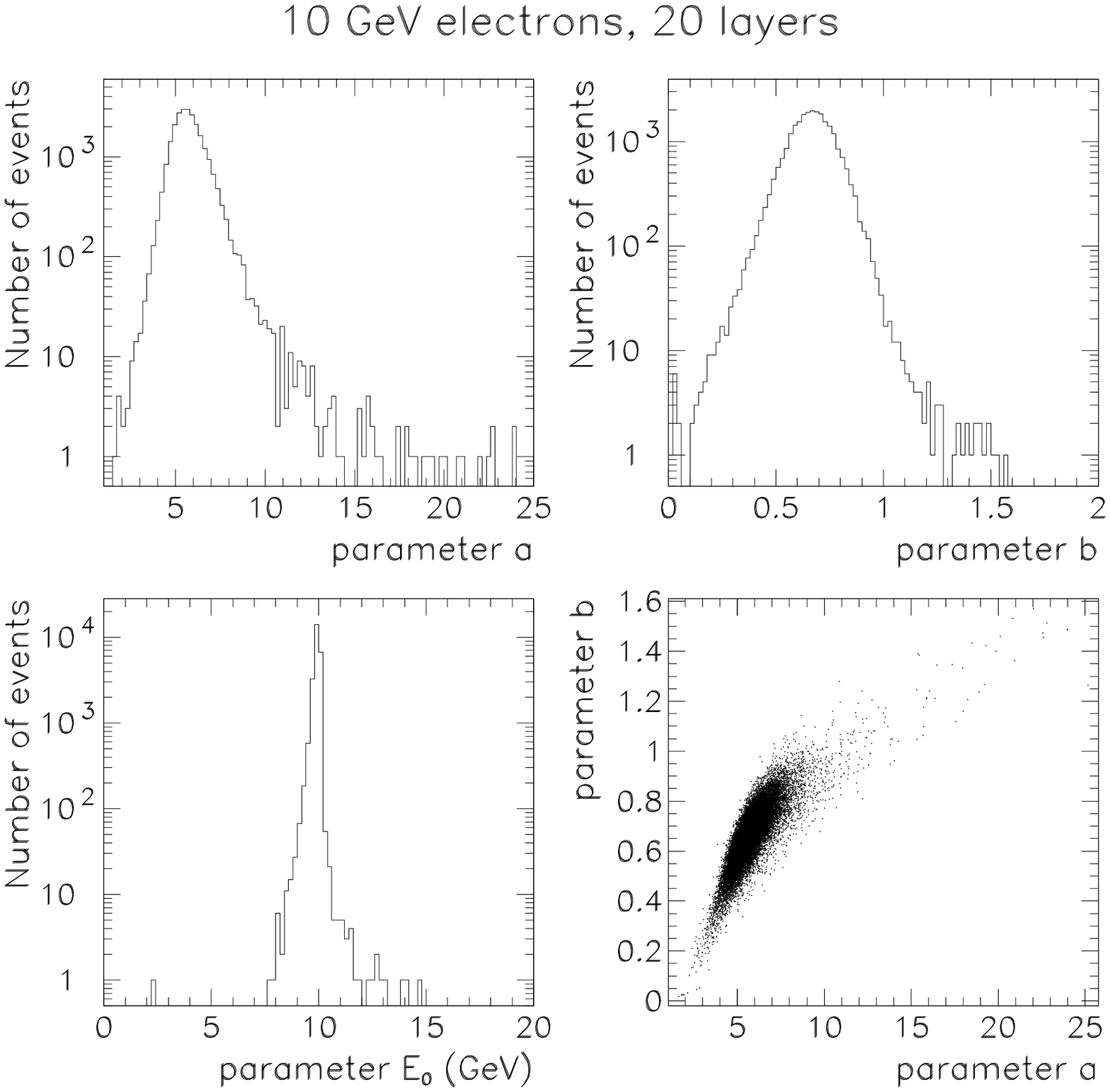}
\caption{Distributions of the parameters obtained from a fit of the $\Gamma$-distribution (see eqn. \ref{eqn:gamma}) to indvidual longitudinal shower profiles. The upper panel shows the case of an 8 layer calorimeter, the lower panel the case of a 20 layer calorimeter. Also shown is the correlation between the $a$ and the $b$ parameter.}
\label{fig:8and20}
\end{figure}

Another method to estimate energy is to fit individual showers with the $\Gamma$ - distribution (see eqn. \ref{eqn:gamma}).
Considering a mono-energetic beam of particles in which all shower maxima are contained within the calorimeter, fitting individual showers will result in Gaussian distributions of $a$,$b$ and $E_0$, centered on their true values. However, as can bee seen in figure \ref{fig:badindfit}, if the shower maximum is not contained within the calorimeter the fit fails badly. Thus, if the energy of the incident particles is high enough such that a significant fraction of showers have maxima outside the calorimeter volume, the distribution of $E_0$ will have significant high energy tails. Figure \ref{fig:8and20} shows in the upper panel the distribution of $E_0$ for a $\sim $9 $X_0$ calorimeter as compared to one which is $\sim$ 22 $X_0$ in depth. For this figure a 10 GeV electron beam was assumed. In the shallower calorimeter about 1 \% of the electrons are estimated to have an energy above 20 GeV. For the deeper calorimeter the situation improves signficantly: the fraction of electrons where the estimated energy is above 11 GeV is only 0.1 \%.

One possible method to account for the high energy tails in the estimation, is to perform a constrained fit, where the parameters a,b and $E_0$ are constrained to ranges determined by Monte Carlo simulations or from data obtained under controlled conditions (beam test).\\

\noindent
In table \ref{tab:simres} we compare the two reconstruction methods presented above with respect to their mean estimated energy and resolution for 1~GeV and 10~GeV photons and electrons. The last layer correction gives a between 40 \% and 50 \% better energy resolution for both particle types. The improvement is larger for the lower energy. 

\begin{table}
\centering
\renewcommand{\arraystretch}{1.5}

\caption{Comparsion of the two energy estimation methods}
\label{tab:simres}
\centering
\begin{tabular}{l|l|l|l|l|l|l|l}
Particle & Energy & \multicolumn{2}{l|}{Energy dep.} & \multicolumn{2}{|l|}{Profile fit}  & \multicolumn{2}{|l}{Last lay.}\\ \cline{3-8}
type     & [GeV] & $E_{tot} $ & $\frac{\sigma_E}{E}$ & $E_{0}$ & $\frac{\sigma_E}{E}$   & $E_{0}$  & $\frac{\sigma_E}{E} $ \\
         &       & [GeV]      & [\%]                 & [GeV]   & [\%]                   & [GeV]    & [\%]             \\
\hline \hline
 e$^-$      & 1  & 0.83 & 6.7    & 0.83 & 10.7 &1.0 & 5.1\\ 
 $\gamma$~\ & 1  & 0.81 & 8.9    & 0.79 & 12.1 &1.0 & 6.6\\
 e$^-$      & 10 & 6.4  & 10.8   & 8.6  & 13.0 &10.1 & 7.6\\ 
 $\gamma$~\ & 10 & 6.2  & 13.3   & 8.2  & 14.8 &10.1 & 9.1\\
\end{tabular}
\end{table}

\subsection{Some remarks on Maximum Likelihood methods}
Using Monte Carlo simulations, in principle, it could be possible to construct the likelihood function:
\begin{equation}
L(E_0|\vec{x})
\end{equation}
where $E_0$ denotes the reconstructed energy and $\vec{x}$ a vector containing all the observations in the calorimeter. In particular, these observations could be the energy deposited in each layer. The parameterizations found in the previous sections could then be used to construct likelihood functions, since they represent probability density functions $P(E_{lay_x}|E_{true})$, i.e. the probability of observing an energy $E_{lay_x}$ given the true energy $E_{true}$. In the simplest case, if all observations are uncorrelated, the multi-dimensional likelihood function will reduce to a multiplication of one-dimensional likelihood functions.\\
In practice, this method is complicated by several factors: firstly there will be an additional parameter and observation (the angle of the incoming photon) and secondly we know that the assumption that the observables are uncorrelated is not correct. Still, one possibility is to test the method under the assumption that correlations can be neglected and see how well it works. In case, correlations can not be neglected the only technically feasible solution (considering limited computing resources) is the application of learning machines, such as artificial neural networks.


\section{Conclusion}
Distributions of energy depositions in an segmented CsI calorimeter have been studied using GEANT4 simulations. Three probability distributions have been fitted to the data: negative binomial, log-normal and Gaussian distributions, none of which gives a good fit over the full shower length. However, negative binomial and log normal distributions give best fits if the shower has a pronounced high energy tail.\\
The performed simulations have also been used to compare two different methods of energy reconstructions. The simplest method is to to add up the energies deposited in each layer of the calorimeter. This method works well for calorimeters whose size is enough such that leakage can be neglected. For finite size calorimeters, a correction for the leakage can be introduced noting that the energy lost due to leakage is correlated with the energy deposited in the last layer.\\ 
Another possible method is to use  a fit of the $\Gamma$ function to individual longitudinal shower profiles. This method works well, if the calorimeter is large enough so that only a non-significant fraction of the showers have their maximum outside the calorimeters sensitive volume. Otherwise, the fit overestimates the energy. Our simulation studies indicate that for 1~Gev and 10 GeV photons and electrons the method of last layer correction gives an improvement in resolution of about 50 \% when compared to the fitting method. The improvement is larger for the 1 GeV case.\\
Finally, we give a short discussion on how the parameterizations of the layer-wise energy depositions presented in this note could be used to improve maximum likelihood estimation methods.


%
%


\section*{Acknowledgment}
The authors would like to thank the K~A~Wallenberg foundation and the Swedish Space Board for financial support.
Computing resources made available by a grant from the G\"oran Gustafsson Foundation are also acknowledged. J.C. acknowledges support from Vetenskapsr{\aa}det, grant nr: 40219601.



%

%

%

\end{document}